\documentclass[conference]{IEEEtran}
\IEEEoverridecommandlockouts
\usepackage{cite}
\usepackage{amsmath,amssymb,amsfonts}
\usepackage{algorithmic}
\usepackage{graphicx}
\usepackage{textcomp}
\usepackage{xcolor}
\usepackage{bm}
\usepackage{url}

\def\BibTeX{{\rm B\kern-.05em{\sc i\kern-.025em b}\kern-.08em
    T\kern-.1667em\lower.7ex\hbox{E}\kern-.125emX}}
\begin{document}

\title{Speech Command Recognition in Computationally Constrained Environments with a Quadratic Self-organized Operational Layer
\thanks{This  work  received  funding  from  the  European Union’s Horizon 2020 research and innovation programme under grant agreement  No  871449  (OpenDR).  This  publication  reflects the  authors’  views  only.  The  European  Commission  is  not responsible for any use that may be made of the information it contains.}
}

\author{\IEEEauthorblockN{Mohammad Soltanian}
\IEEEauthorblockA{\textit{Department of Computing Sciences} \\
\textit{Tampere University}\\
Tampere, Finland \\
soltanianmm@gmail.com}
\and
\IEEEauthorblockN{Junaid Malik}
\IEEEauthorblockA{\textit{Department of Computing Sciences} \\
\textit{Tampere University}\\
Tampere, Finland \\
junaid.malik@tuni.fi}
\and
\IEEEauthorblockN{Jenni Raitoharju}
\IEEEauthorblockA{\textit{Programme for Environmental Information} \\
\textit{Finnish Environment Institute}\\
Jyv\"askyl\"a, Finland \\
jenni.raitoharju@syke.fi}
\and
\IEEEauthorblockN{Alexandros Iosifidis}
\IEEEauthorblockA{\textit{Department of Engineering} \\
\textit{Aarhus University}\\
Aarhus, Denmark \\
ai@ece.au.dk}
\and
\IEEEauthorblockN{Serkan Kiranyaz}
\IEEEauthorblockA{\textit{Electrical Engineering Department} \\
\textit{Qatar University}\\
Doha, Qatar \\
mkiranyaz@qu.edu.qa}
\and
\IEEEauthorblockN{Moncef Gabbouj}
\IEEEauthorblockA{\textit{Department of Computing Sciences} \\
\textit{Tampere University}\\
Tampere, Finland \\
moncef.gabbouj@tuni.fi}
}

\maketitle

\begin{abstract}
Automatic classification of speech commands has revolutionized human computer interactions in robotic applications. However, employed recognition models usually follow the methodology of deep learning with complicated networks which are memory and energy hungry. So, there is a need to either squeeze these complicated models or use more efficient light-weight models in order to be able to implement the resulting classifiers on embedded devices. In this paper, we pick the second approach and propose a network layer to enhance the speech command recognition capability of a lightweight network and demonstrate the result via experiments. The employed method borrows the ideas of Taylor expansion and quadratic forms to construct a better representation of features in both input and hidden layers. This richer representation results in recognition accuracy improvement as shown by extensive experiments on Google speech commands (GSC) and synthetic speech commands (SSC) datasets.
\end{abstract}

\section{Introduction}
\label{sec:intro}

Automatic Speech recognition (ASR) is the art and science of automatic identification of speech. Speech commands recognition (SCR) is a subcategory of ASR in which the machine identifies short spoken commands. It is a key technology which enables machines to understand human commands. Possible applications include smart-phones with mobile assistants, robots which follow human spoken instructions, and smart home assistants, like Amazon Echo and Google Home. These kinds of applications are increasingly affecting human machine interactions. There
are also many use cases for this technology in both medicine and education, e.g., for blind or handicapped people \cite{1998Single-surgeonThoracoscopicSurgeryWith,2014EducationalSystemForThe}.
As many of these applications rely on small sized devices which have limited computational capacities, fast and lightweight implementations are needed to work smoothly in real time \cite{2018FPGAImplementationOfConvolutional}.

Classical SCR approaches rely on extracting discriminative features from the raw speech. Frequency domain features like fast Fourier transform are among the most widely used features \cite{1989Speaker-independentPhoneRecognitionUsing,2013TechniquesForFeatureExtraction} to convert the time domain speech into frequency space. After feature extraction, an acoustic model such as hidden Markov model (HMM) was classically used to represent the sequence of words of phonemes \cite{2014SpeechAndLanguageProcessing.}. Combination of neural networks and HMM-based approaches have also been in use since more than 20 years ago \cite{2012ConnectionistSpeechRecognition:A,1986TheTRACEModelOf}. 

Recently, deep learning-based approaches have dominated, in terms of performance accuracy, over conventional machine learning methods for many different applications \cite{2012DeepNeuralNetworksFor,2015DeepLearning,2016DeepLearning}. Neural networks like long short term memory (LSTM) have raised the accuracy of natural language processing systems beyond the traditional approaches \cite{1999LearningToForget:Continual,2016LSTM:ASearchSpace}. Also, the neural network based approaches have dominated the whole SCR history and have been accepted as the state of the art tool by the speech recognition society \cite{2014Small-footprintKeywordSpottingUsing,2017HelloEdge:KeywordSpotting,2018DeepResidualLearningFor,2019TemporalConvolutionForReal-time}. In the meanwhile, several different neural network architectures like convolutional neural networks (CNNs), LSTMs, and gated recurrent units (GRUs) have been successfully applied to SCR \cite{2017HelloEdge:KeywordSpotting,2018DeepResidualLearningFor,2019TemporalConvolutionForReal-time}. Direct application of CNNs on the raw audio have been explored in \cite{2018SpeakerRecognitionFromRaw,2018InterpretableConvolutionalFiltersWith,2018End-to-endSpeechRecognitionFrom}. This is while the low computational and memory capacities of the embedded devices are still serious barriers to fully exploit the benefits of highly accurate complicated deep networks.

To tackle the problem of computational complexity on embedded devices, there are usually two general approaches. One can squeeze the state of the art deep models to reduce the computational costs without severely degrading the performance accuracy \cite{2019AnOptimizedRecurrentUnit,2020Small-FootprintOpen-VocabularyKeywordSpotting}. Alternatively, more efficient deep networks and methods can be employed from the beginning in order to achieve better accuracy rates with lower computational costs \cite{2019OperationalNeuralNetworks,2020Self-OrganizedOperationalNeuralNetworks}. The second approach especially becomes important when training examples are scarce or the input signals are of low quality or resolution. Using either of these ways, one can implement a reasonable classifier on energy constrained embedded devices.

There are many studies focusing on squeezing the complicated networks or building lightweight networks from the beginning. Progressive operational perceptrons (POPs)\cite{2017ProgressiveOperationalPerceptrons} try to generalize the conventional multilayer perceptrons (MLPs) using generalized operational perceptrons (GOPs). GOPs have a distinct set of operators, to mimic the synaptic connections of biological neurons, which enables them to better learn the hidden semantics of data. PyGOP \cite{2019PyGOP:APythonLibrary} is a library for implementation of GOPs in both CPU and GPU. Authors in \cite{2020ProgressiveOperationalPerceptronsWith} accelerate the learning process in POPs employing the concept of memory and information from previously learned layers. The idea of biologically inspired neurons is further explored in \cite{2019HeterogeneousMultilayerGeneralizedOperational}, where each neuron in the network can have its distinct set of operators, which results in a compact network.

In \cite{2018ImprovingEfficiencyInConvolutional}, a generic neural network layer structure employing a multi-linear projection is proposed, which requires several times less memory compared to traditional CNNs, while achieving better performance. Authors in \cite{2017HelloEdge:KeywordSpotting} used deep networks with dilated convolutions to achieve a high accuracy with lower number of parameters. 
SincNet Kernels \cite{2018InterpretableConvolutionalFiltersWith} at the first convolutional layer of a CNN is another method to shrink the CNN by lowering the number of learnable parameters. This method is used for speaker recognition in \cite{2018SpeakerRecognitionFromRaw} and later for phoneme recognition \cite{2018InterpretableConvolutionalFiltersWith}. Depth-wise separable convolutions (DSCconvs) have also been introduced firstly in image processing field \cite{2017Xception:DeepLearningWith,2017Mobilenets:EfficientConvolutionalNeural} and later in SCR, keyword spotting \cite{2017HelloEdge:KeywordSpotting}, and machine translation \cite{2017DepthwiseSeparableConvolutionsFor} in order to reduce the amount of required computations. 

In this paper, we propose a new network layer to enhance the speech command recognition capability using a lightweight network. To this end, we use a novel combination of self-organized operational neural networks (SelfONNs) \cite{2020Self-OrganizedOperationalNeuralNetworks} and quadratic forms kernels \cite{2017Non-linearConvolutionFiltersFor}. The experiments on Google speech commands (GSC) \cite{2018SpeechCommands:ADataset} and synthetic speech commands (SSC) datasets\footnote{\url{https://kaggle.com/jbuchner/synthetic-speech-commands-dataset}} show the effectiveness of the proposed approach for recognition of speech commands especially in lightweight networks.
To the best of our knowledge, we are the first to propose a combination of SelfONNs and quadratic kernels to make a new network layer type to boost the performance of a lightweight CNN. We are also the first to apply concepts of SelfONN and quadratic kernels on the task of SCR.

In the following, we describe the most relevant work in Section~II. The proposed quadratic SelfONN approach to the problem of SCR is introduced in Section~III, and the experiments are discussed in Section~IV. Finally, the paper is concluded in Section~V.

\section{Related Work}

\subsection{Self-organized Operational Neural Networks}
Operational  Neural  Networks  (ONNs)  \cite{2019OperationalNeuralNetworks} are heterogeneous  networks  with  a  generalized neuron  model that includes a dictionary of nodal operators (instead of just an ordinary convolution) to boost the performance of the conventional CNNs. However, they need a greedy search to find the optimal operators which result in the best accuracy. The performance is also highly dependent on the selection of the initial operator set dictionary. In order to resolve these issues, SelfONNs with generative neurons were recently proposed in \cite{2020Self-OrganizedOperationalNeuralNetworks}, which are able to  adapt  the  nodal  operators  of  each neuron during the training phase. This removes the need for a fixed operator set and greedy search within that library to find the optimal set of operators. The authors show, via extensive experiments, the superiority of SelfONNs over ONNs in four applications: image synthesis, image denoising, face segmentation, and image transformation.

Let us denote by $\bm{X} \in \mathbb{R}^{H \times W \times C_{in}}$ the input tensor to a layer and by $\bm{X}_{(i,j)} \in \mathbb{R}^{h \times w \times C_{in}}$ a sub-tensor of it centered at the position $(i,j)$. Let us also denote by $\bm{W}_{c_{out}} \in \mathbb{R}^{h \times w \times C_{in}}, \:c_{out}=1,\dots,C_{out}$ the $c_{out}$-th filter of the layer. Convolutional neurons in CNNs convolve $\bm{X}$ with $\bm{W}_{c_{out}}$ and add an offset, which corresponds to applying the following calculation for every position $(i,j)$ of the input tensor $\bm{X}$:
\begin{align}
    &\bm{Y}_{c_{out}}(i,j) \nonumber \\
    & = \sum_{k,m,c_{in}=1}^{h,w,C_{in}} \bm{W}_{c_{out}}(k,m,c_{in}) \bm{X}_{(i,j)}(k,m,c_{in})  +b_{c_{out}} \nonumber \\
    & = \bm{w}_{c_{out}}^T \bm{x}_{(i,j)} + b_{c_{out}}, \label{Eq:CNNlayer}
\end{align}
where $\bm{Y}_{c_{out}}(i,j)$ is the $(i,j)$-th element of the output feature map $\bm{Y}_{c_{out}}$, $b_{c_{out}}$ is the bias term, and $\bm{x}_{(i,j)}$ and $\bm{w}_{c_{out}}$ are the vectorized versions of $\bm{X}_{(i,j)}$ and $\bm{W}_{c_{out}}$, respectively. 
The feature maps $\bm{Y}_{c_{out}}, \: c_{out} = 1,\dots,C_{out}$ are concatenated to form the tensor\footnote{We assume that appropriate zero-padding is used.} $\bm{Y} \in \mathbb{R}^{H \times W \times C_{out}}$, and an element-wise activation function is applied to produce the output of the layer. 

In ONNs, the transformation in Eq. (\ref{Eq:CNNlayer}) is generalized to the following:
\begin{equation}
\bm{Y}_{c_{out}}(i,j) = \Psi(\bm{x}_{(i,j)},\bm{w}_{c_{out}}) + b_{c_{out}},
\end{equation}
where $\Psi(\cdot)$ is an arbitrary (differentiable) nodal function and may be a combination of different functions. The selection of $\Psi(\cdot)$ is conducted by following a search strategy during training, which leads to a slow training process. When $\Psi(\cdot)$ is determined to be the dot-product of its arguments, the ONN layer is equal to an ordinary convolutional layer.

In SelfONNs, instead of searching for an optimal function $\Psi(\cdot)$ during training, a suitable function is approximated with a truncated Taylor series expansion:
\begin{align}
    &\Psi(\bm{x}_{(i,j)},\bm{w}_{c_{out},1}, \dots, \bm{w}_{c_{out},Q}) = \bm{w}_{c_{out},0}^T \bm{1} + \bm{w}_{c_{out},1}^T \bm{x}_{(i,j)} \nonumber \\&+ \bm{w}_{c_{out},2}^T \bm{x}_{(i,j)}^2 +  
     ... + \bm{w}_{c_{out},Q}^T \bm{x}_{(i,j)}^Q = \sum_{q=0}^Q \bm{w}_{c_{out},q}^T \bm{x}_{(i,j)}^q, \label{Eq:TaylorExpansion}
\end{align}
where $\bm{1}$ is a vectors of '1's, in $\bm{x}_{(i,j)}^q$ the power is applied in an element-wise manner and $\bm{w}_{c_{out},q}$ are the learnable weights interacting with $\bm{x}_{(i,j)}^q$. Thus, each neuron of SelfONN performs the following transformation:
\begin{equation}
\bm{Y}_{c_{out}}(i,j) = \sum_{q=1}^Q \bm{w}_{c_{out},q}^T \bm{x}_{(i,j)}^q + b_{c_{out}},
\end{equation}
where the term $\bm{w}_{c_{out},0}$ is absorbed in the learnable bias $b_{c_{out}}$. As the goal is not to estimate any particular function $\Psi(\cdot)$, but to learn the most suitable nodal function for the learning problem at hand, $\bm{w}_{c_{out},q}, \:q=1,\dots, Q$ are learned via gradient-based optimization.

\subsection{Quadratic-form Kernels}
The core idea behind the network layer based on the quadratic form expansion \cite{2017Non-linearConvolutionFiltersFor} is that it generalizes the linear convolution by taking into account the cross correlation between the input elements within the receptive field of the layer kernel, i.e.:
\begin{equation}
    \bm{Y}_{c_{out}}(i,j) = \bm{x}_{(i,j)}^T \bm{\Omega}_{c_{out}} \bm{x}_{(i,j)} + \bm{w}_{c_{out}}^T\bm{x}_{(i,j)} + b_{c_{out}}.  \label{Eq:QuadraticConvolution}
\end{equation}
Because the quadratic term is performed in a channel-wise manner, $\bm{\Omega}_{c_{out}}$ is a block-diagonal matrix formed by upper-triangular blocks. The diagonal block interacting with the elements of $\bm{X}$ in its $c_{in}$-th channel, i.e. $\bm{\Omega}_{c_{out}}^{(c_{in},c_{in})}$ is a Volterra kernel of the form:
\begin{equation}
\bm{\Omega}_{c_{out}}^{(c_{in},c_{in})} =  \begin{pmatrix}
  \omega_{1,1} & \omega_{1,2} & \cdots & \omega_{1,n} \\
  0 & \omega_{2,2} & \cdots & \omega_{2,n} \\
  \vdots  & \vdots  & \ddots & \vdots  \\
  0 & 0 & \cdots & \omega_{n,n} 
 \end{pmatrix}, \label{Eq:VolterraKernel}
\end{equation}
where $n = hw$.

\section{Proposed Quadratic Self-organized Operational Neural Network method}
\begin{figure*}[htb]
	\centerline{\includegraphics[trim=280 400 250 180, width=0.8\paperwidth]{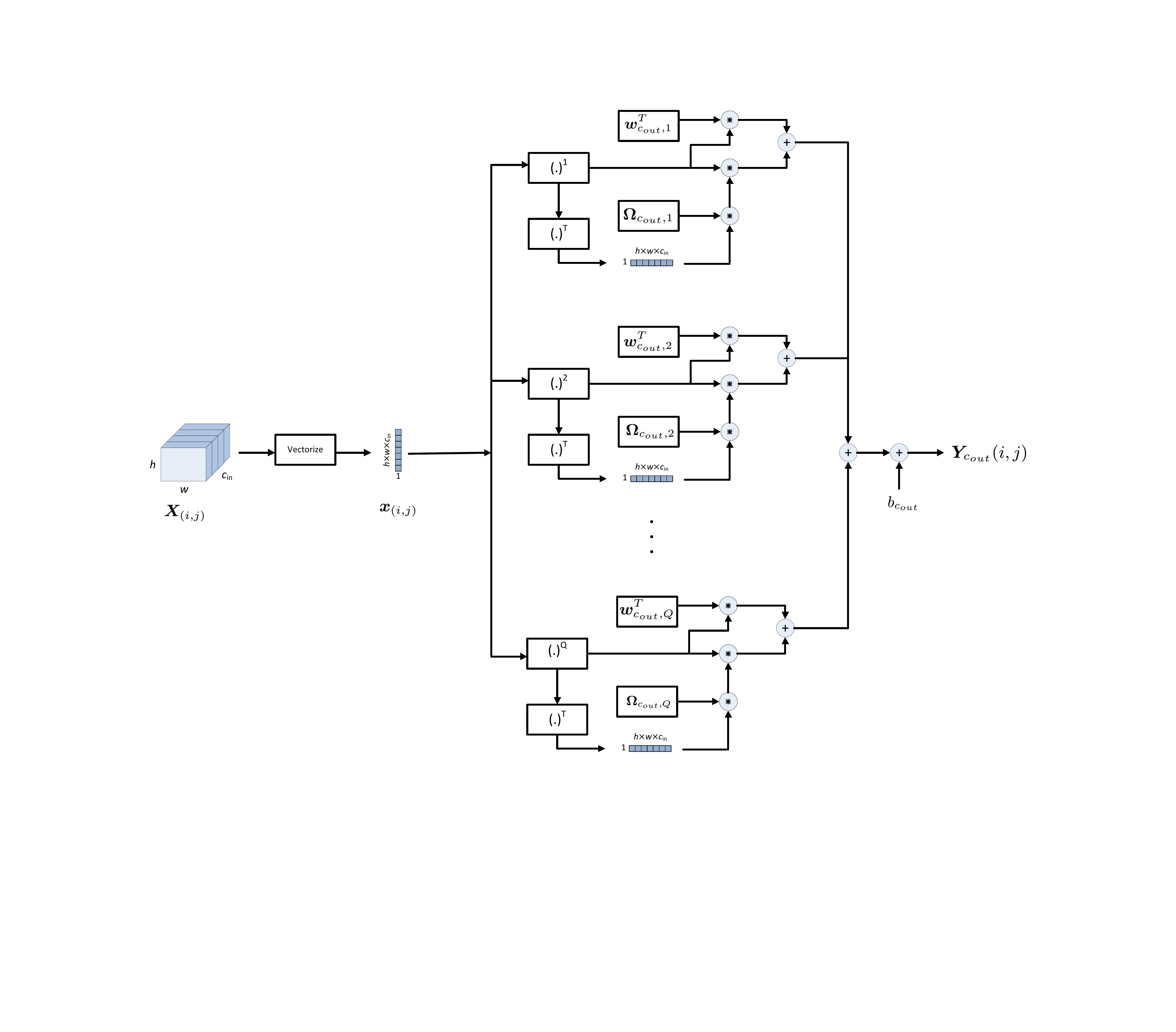}}
	\caption{Flow diagram of the proposed quadratic-SelfONN based layer: The spectrogram, MFCC, or any feature map is fed as the input. The input is then expanded as a power series, and linear and quadratic convolutions are applied. The results are summed together and the bias is added.}
	\label{fig:flow}
\end{figure*}

The concepts of SelfONN and quadratic convolution have not yet been applied to SCR problem. We combine these two layer types into a new type of layer which inherits advantages of both SelfONN (by a generalized convolution) and quadratic kernels (by modeling cross correlation of input feature points). To do so, we merge the Taylor series expansion in Eq. (\ref{Eq:TaylorExpansion}) with the quadratic-form convolution in Eq. (\ref{Eq:QuadraticConvolution}) as follows:
\begin{align}
&\bm{Y}_{c_{out}}(i,j) \nonumber \\ 
& = \sum_{q=1}^Q (\bm{x}_{(i,j)}^q)^T \bm{\Omega}_{c_{out},q} \bm{x}_{(i,j)}^q + \sum_{q=1}^Q \bm{w}_{c_{out},q}^T \bm{x}_{(i,j)}^q + b_{c_{out}}. 
\label{eq:combination}
\end{align}
where two sets of learnable weights $\bm{\Omega}_{c_{out},q}$ and $\bm{w}_{c_{out},q}$ are interacting with $\bm{x}_{(i,j)}^q$. The matrix $\bm{\Omega}_{c_{out},q}$ in Eq. (\ref{eq:combination}) is again a block-diagonal matrix as in Eq. (\ref{Eq:QuadraticConvolution}) but, instead of using upper-triangular blocks, all values of its diagonal blocks are learnable.
The flow diagram of the proposed layer is shown in Figure~\ref{fig:flow}. 

\section{Experiments}
\subsection{Datasets}
The proposed model is evaluated on Google Speech Commands (GSC)  \cite{2018SpeechCommands:ADataset} and Synthetic Speech Commands (SSC) \footnote{\url{https://kaggle.com/jbuchner/synthetic-speech-commands-dataset}}  datasets. GSC is the most widely used benchmark for SCR systems. The training set of it consists of more than 60,000 audio files categorized into 32 directories with the folder names being the label of the audio clips. A subset of the dataset, which is commonly used, e.g., in Kaggle competition, contains "on", "off", "yes", "no", "left", "right", "up", "down", "stop", and "go" commands. This is the subset we use for our evaluations in the current work. Each audio file in GSC is a voice clip with length of almost 1 second. Every audio file is in the form of 16-bit little-endian PCM-encoded WAVE file at a sampling frequency of 16000 Hz.

SSC is automatically generated and is actually text-to-speech counterpart for the GSC. It contains two versions, the normal version and very noisy version, each of which consists of 41849 audio files categorized into 30 classes. Since GSC is a low noise dataset, the very noisy version of SSC is employed in the experiments to assess the performance of the proposed method on noisy data as well. The same subset of classes as for GSC is used SSC in our experiments. In contrary to GSC, the audio files in SSC are not segmented into different train, test, and validation subsets. So, we randomly partition each category into train (80 \%), test (10 \%), and validation (10 \%).

The accuracy of the proposed method against the baseline is evaluated based on the number of correctly predicted labels in the test set divided by the total number of test samples, for both GSC and SSC datasets.

\subsection{Data Pre-processing}
The 1D raw input audio is converted to a 2D signal before being fed to the network. Actually, each input speech wave is segmented to 30 milliseconds windows with 10 milliseconds overlap. Then, Mel-frequency cepstral coefficients (MFCCs) are extracted from each window. Since the configuration is aimed to be used in computationally constrained environments, only 20 cepstral coefficients are kept at each window. Although 40 or more cepstral coefficients are usually used in research papers like \cite{2018DeepResidualLearningFor,2017HelloEdge:KeywordSpotting}, it is widely believed that 15 or more cepstral coefficients will preserve the essential information in audio signals for classification applications.
%The number of frequency bins equals to $0.5*\text{window size} * \text{sampling frequecy} = 160$. The magnitude of STFT is then converted to logarithmic scale and down-sampled with different rates to simulate spectrogram with different temporal and frequency resolutions.
The resulting single channel 2D MFCC $\in \mathbb{R}^{20 \times 51}$ is finally normalized to lie between -1 and 1.

\subsection{Network Structure}
For comparison of the proposed layer with the baseline convolutional layer in terms of accuracy of final speech commands classification, we need to use the layers in a deep structure which is end-to-end trainable. Since we aim to use the resulting network in computationally constrained environments, like embedded devices in robotic applications, we use a lightweight basic model for the evaluations. Indeed, the model could not dominate the complex state-of-the-art networks, but it is quite enough for a proof of concept, i.e., showing superiority of the proposed layer over the baseline convolutional layer. So, we choose a structure which is similar to LeNet-1 \cite{1998Gradient-basedLearningAppliedToa}.  The original LeNet-1 has two convolutional and one fully connected layers. The network employed here has the same structure, but the convolutional layers are replaced with the proposed quadratic SelfONN layer. The codes will also be made freely available on-line upon publication of the paper.

After each convolutional layer a max pooling layer and a tangent hyperbolic activation are used. Dropout is also used to prevent over-fitting. The proposed layer and the resulting network are implemented in PyTorch v1.4.0. 

\subsection{Hyper-parameters}
For training of the network, the accuracy over validation set is computed at each epoch and the training proceeds if both following conditions are met: the epoch number is less than the maximum allowable epochs (100 in our simulations) and the accuracy over validation set is not less than the maximum achieved validation accuracy for the last 10 epochs. The mini-batch size is set to 50. Based on cross  validation  on  the  training  set  using  the  CNN comprised of the baseline convolutional layers, the kernel size is set to $3 \times 3$, and dilation and stride to 1. Padding of size 2 is also applied to the all sides of the input of the convolutional layers. The number of neurons (or channels) in two convolutional layers and one fully connected layer equals to 20, 20, and 10, respectively. 
Stochastic gradient descent with momentum = 0.9 and learning rate = 0.01 is used for network training.

\subsection{Experimental Results}
%Figure \ref{fig:sim1} shows the convergence curves in the training process of the CNN for both convolutional and proposed layers at two different resolutions, one fourth of the original spectrogram resolution, and the other one eights of the original spectrogram resolution. As it could be seen, in both resolutions the proposed layer totally outperforms the convolutional one.
%
%\begin{figure}[htbp]
%	\centerline{\includegraphics[width=\columnwidth]{figs/sim1.eps}}
%	\caption{Comparison of the convolutional and proposed layer in terms of validation loss during training for SCR task.}
%	\label{fig:sim1}
%\end{figure}

\begin{figure}[htb]
	\centerline{\includegraphics[width=1.1\columnwidth]{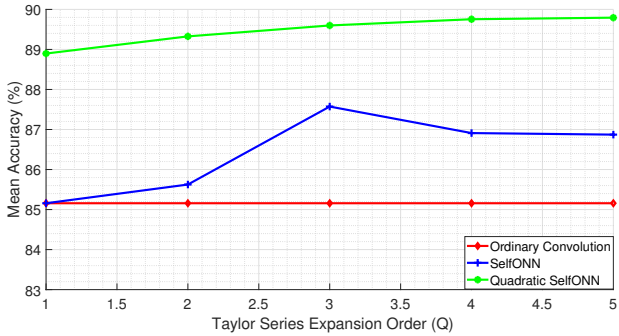}}
	\caption{Comparison of the convolutional, SelfONN, and proposed layer in terms of test accuracy for SCR task over GSC dataset.}
	\label{fig:sim1}
\end{figure}

\begin{figure}[htb]
	\centerline{\includegraphics[width=1.1\columnwidth]{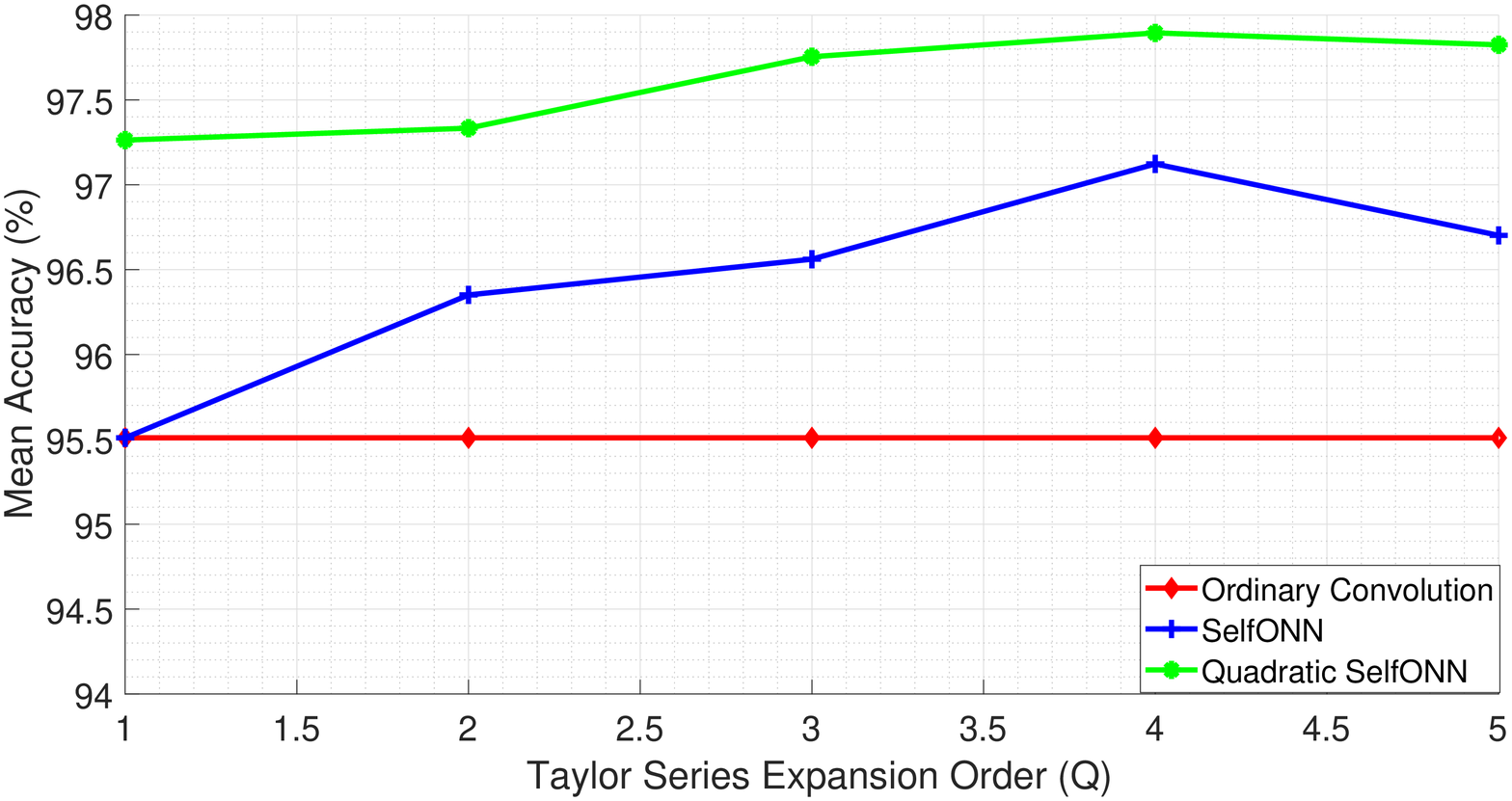}}
	\caption{Comparison of the convolutional, SelfONN, and proposed layer in terms of test accuracy for SCR task over SSC dataset.}
	\label{fig:sim2}
\end{figure}

Figures \ref{fig:sim1} and \ref{fig:sim2} show the comparison of accuracy versus maximum Taylor series power ($Q$) over the test set between convolutional, SelfONN, and proposed layers for GSC and SSC datasets, respectively. As shown, the network comprised of the proposed layer outperforms the network with convolutional or SelfONN layers in terms of test accuracy over both datasets. For SelfONN, the values of $Q=3$ and $Q=4$ result in the best accuracy for GSC and SSC datasets, respectively. The accuracy of the proposed quadratic SelfONN layer first increases with increase in $Q$, but the accuracy is saturated when $Q$ further increases.

As can be inferred from Figures~\ref{fig:sim1}~and~\ref{fig:sim2}, which is also summarized in Table~\ref{table:table1}, the accuracy gains for the proposed layer over the baseline convolutional layer are nearly 4.6\% and 2.2\% for GSC and SSC datasets, respectively. Similarly, the accuracy gains of the proposed layer over SelfONN layer are nearly 2.4\% and 0.8\% for GSC and SSC datasets, respectively.

\begin{table}[!t]
	\renewcommand{\arraystretch}{1.3}
	\caption{Performance comparison of the proposed layer with ordinary convolutional and SelfONN layers over GSC and SSC datasets.}
	\label{table:table1}
	\centering
	\begin{tabular}{|c|c|c|c|}
		\hline
		Dataset & Ordinary convolution & SelfONN & Proposed\\
		\hline
		\hline
		GSC & 85.2 & 87.6 & 89.8\\
		\hline
		SSC & 95.5 & 97.1 & 97.9\\
		\hline
	\end{tabular}
\end{table}

\begin{figure}[htb]
	\centerline{\includegraphics[width=1.1\columnwidth]{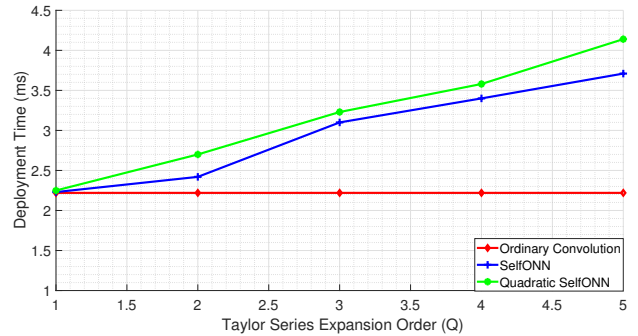}}
	\caption{Comparison of the convolutional, SelfONN, and the proposed layers in terms of deployment time on a Nvidia GeForce GTX 1080 for SCR task over GSC dataset.}
	\label{fig:sim3}
\end{figure}

Figure~\ref{fig:sim3} evaluates the proposed layer based on the deployment complexity. To this end, the deployment time for a single 1 second speech command is computed by averaging the deployment time over the whole test set. The simulations and learning process are run on a machine with a Nvidia GeForce GTX 1080.

As can be seen, the deployment times of the proposed layer are higher than those of the convolution layer. However, the deployment time of the proposed layer is quite close to that of SelfONN.

\subsection{Comparison to the State-of-the-art}
According to leader board section of Kaggle challenge on GSC dataset\footnote{\url{https://kaggle.com/c/tensorflow-speech-recognition-challenge}} , the first team has achieved an accuracy around 91\%. However, they probably used much more complex networks or ensemble of classifiers. They also used the original wave signals with full resolution spectrograms or higher number of cepstral coefficients. Since, the purpose of our work is the proof of concept of the proposed layer in computationally limited environments, we just used a very light-weight network, with small resolution inputs and compared the accuracy with the baseline convolutional layers. Even with these limitations, the achieved accuracy is quite high. However, it is expected that employing the proposed layer in deeper networks will result in higher accuracy values. A similar discussion also holds for SSC dataset.

\section{Conclusion}

An extension of the ordinary convolutional layer is proposed which is highly efficient especially at computationally constrained environments. It is especially appealing for robotic applications, where the resource limitations will not allow to use highly deep structures. The performance gap is promising with the proposed layer for embedded SCR applications. Future work will involve using the proposed layer in more complex networks in combination with network squeezing methods like network pruning, distillation, and sparsification of the fully connected layers.

%\begin{table}[htbp]
%\caption{Table Type Styles}
%\begin{center}
%\begin{tabular}{|c|c|c|c|}
%\hline
%\textbf{Table}&\multicolumn{3}{|c|}{\textbf{Table Column Head}} \\
%\cline{2-4} 
%\textbf{Head} & \textbf{\textit{Table column subhead}}& \textbf{\textit{Subhead}}& \textbf{\textit{Subhead}} \\
%\hline
%copy& More table copy$^{\mathrm{a}}$& &  \\
%\hline
%\multicolumn{4}{l}{$^{\mathrm{a}}$Sample of a Table footnote.}
%\end{tabular}
%\label{tab1}
%\end{center}
%\end{table}

\bibliographystyle{IEEEbib}
\bibliography{SpeechRecognition}

\end{document}